\documentclass[preprint2]{aastex}

\shorttitle{Radio Stars}
\shortauthors{Kimball et al.}

\begin{document}
\bibliographystyle{apj}

\title{A Sample of Candidate Radio Stars in FIRST and SDSS}
\author{Amy E. Kimball\altaffilmark{1}}
\affil{akimball@astro.washington.edu}
\author{Gillian R. Knapp\altaffilmark{2}}
\author{\v{Z}eljko Ivezi\'{c}\altaffilmark{1}}
\author{Andrew A. West\altaffilmark{3}}
\author{John J. Bochanski\altaffilmark{3}}
\author{Richard M. Plotkin\altaffilmark{1}}
\author{Michael S. Gordon\altaffilmark{2}}

\altaffiltext{1}{Department of Astronomy, University of Washington, Box 351580,
  Seattle, WA 98195-1580}
\altaffiltext{2}{Department of Astrophysical Sciences, Princeton University,
  Princeton, NJ 08544} 
\altaffiltext{3}{Kavli Institute for Astrophysics and Space Research,
  Massachusetts Institute of Technology, Cambridge, MA 02139} 

\begin{abstract}
We conduct a search for radio stars by combining radio and optical data from
the Faint Images of the Radio Sky at Twenty cm survey (FIRST) and the Sloan 
Digital Sky Survey (SDSS).  The faint limit of SDSS makes possible a
homogeneous search for radio emission from stars of low optical luminosity.
We select a sample of 112 candidate radio stars in the magnitude range
$15<i\lesssim19.1$ and with radio flux $S_{20}\geq1.25$mJy, from about 7000
deg$^2$ of sky.  The selection criteria are positional coincidence within
$1\arcsec$, radio and optical point source morphology, and an SDSS spectrum
classified as stellar.  The sample contamination is estimated by random
matching to be $108\pm13$, suggesting that at most a small fraction of the
selected candidates are genuine radio stars.  Therefore, we rule out a very
rare population of extremely radio-loud stars: no more than 1.2 of every
million stars in the magnitude range $15<i<19.1$ stars has radio flux
$S_{20}\geq1.25$ mJy.  We investigate the optical and radio colors of the
sample to find candidates that show the largest likelihood of being real radio
stars.  The significant outliers from the stellar locus, as well as the
magnetically active stars, are the best candidates for follow-up radio
observations.  We conclude that, while the present wide-area radio surveys are
not sensitive enough to provide homogeneous samples of the extremely rare radio
stars, upcoming surveys which exploit the great sensitivity of current and
planned telescopes do have sufficient sensitivity and will allow the properties
of this class of object to be investigated in detail. 
\end{abstract}


\section{INTRODUCTION}
\label{sec:intro}

The light from stars dominates the optical sky, while the radio sky's
contribution from stars is very small.  However, significant radio 
emission has been detected from \emph{active} stars in the form of synchrotron, 
gyrosynchrotron, or electron cyclotron maser emission \citep{dulk,gudel}.
Some of the non-thermal processes that lead to these types of emission---
plasma heating and particle acceleration in stellar coronae--- are seen in our
own Sun, but the relevant energies for active stars are much larger.  Radio
emission at the relative level of that emitted by the Sun remains undetected
from even the closest solar-type main sequence stars to the present
day.\footnote{Typical radio flux from a star identical to the Sun, and with
  apparent magnitude $m_V=20$, would reach about $3.0\times10^{-6}$ mJy.}
The quiescent, slowly varying radio emission seen in many active stars
\citep[e.g.][and references therein]{gudel} has no solar counterpart.

With the great increase in the sensitivity of radio surveys in the last several
decades, along with the more accurate source positions allowed by radio
interferometry, both thermal and non-thermal radio emission have now been
detected from hundreds of stars of many different types
\citep{hjellming1986,wendker87,altenhoff,wendker95,radiostars96}.  
These include pre-main-sequence stars \citep[T Tau and Herbig
  Ae/Be\footnote{The ``e'' following the spectral class indicates emission in
    the spectrum.} stars;][]{guedel1989,white1992,skinner1993},
rapidly-rotating main-sequence stars \citep{lim1995,berger02}, X-ray bright
main sequence stars \citep{guedel1995}, magnetic stars
\citep{drake1987a,leone1996,berger06,berger08}, cool giants with extended
chromospheres and photospheres
\citep{newell1982,drake1986,drake1987,knapp1995,reid1997}, OB stars with winds
\citep{bieging1989,drake1990,phillips1990}, Wolf-Rayet stars
\citep{chapman1999}, dMe flare stars \citep{White1989,osten06}, and various
classes of interacting binaries and cataclysmic variables.  The radio radiation
from these stars is very faint, at the few mJy level.

The first large unbiased study of radio stars ($\sim5000\deg^2$ of sky at high
galactic latitude) was performed by \citet[][hereafter H99]{helfand1999}, who
compared the Faint Images of the Radio Sky at Twenty cm survey
\citep[FIRST;][]{first} to several catalogs of bright stars with high
astrometric precision: the Hipparcos 
catalog \citep{hipparcos}, the Tycho catalog \citep{tycho}, the Guide Star
Catalog \citep{gsc1}, and stars within 25 pc of the  
Sun.  They emphasize the need for accurate positions: the rarity of radio stars
to the FIRST flux limit ($\sim$ 1 mJy), combined with the high density of faint
extragalactic radio sources, ensures random matches between stellar and radio
sources in sufficient numbers to confuse the cataloging of true radio stars,
unless both radio and optical positions are known to better than $1\arcsec$.
H99 identified 26 radio stars in their study, about one per $190\deg^2$, and
showed that the fraction of stars with radio emission above the FIRST limit
declines steeply with optical magnitude to $m_V\gtrsim15$.  The fraction of
radio stars at fainter magnitudes is unknown.  \citet[][hereafter
  KI08]{kimball2008} searched for radio stars in a combined radio---optical
catalog with observations from FIRST and from the optical Sloan Digital Sky
Survey (SDSS).  Quasars are the most common radio source above flux densities
of a few mJy; therefore a sample of optical point sources with radio emission
is likely to be srongly dominated by quasars.  KI08 approached this problem by
applying a conservative photometric color cut, using the fact that quasars and
stars lie in different locations in SDSS optical color-color diagrams
\citep{richards01}.  The sample was limited to sources sufficiently bright to
be included in the SDSS quasar spectroscopic target selection
\citep[$i<19.1$,][]{dr5quasars}.  Only 20\% of the photometrically-selected
candidate radio stars actually showed stellar spectra, while the rest were
quasars with stellar-like colors.  KI08 concluded that simple color critera are
not sufficient to select a clean sample of radio stars, and that spectroscopic
observations are necessary to distinguish between quasars and stars.

In this paper we continue the search for radio stars in the SDSS using a sample
with spectroscopic identifications.  We present 112 candidates selected by
matching FIRST detections and SDSS point sources within 1 arcsec.  The sample
comprises sources brighter than $r=20.5$ in the optical and 1.25\,mJy at
20\,cm, with SDSS spectra classified as stellar both by the automated reduction
pipelines and visually.  The SDSS spectroscopic targeting implicitly imposes
soft magnitude limits of $15<i\lesssim19.1$.  In this magnitude range,
approximately 1\% of SDSS stars have spectroscopic data.  However, \emph{all}
objects in this range which are close to a FIRST source are targeted for
spectra \citep{edr}, so the completeness of SDSS radio---optical sources is
well understood.

We are searching in a different region of the radio---optical parameter space
from H99, who also matched to FIRST but used optical catalogs brighter than
$m_v=15$.  The SDSS has a saturation limit of $m_i=15$.  By extending to
several optical magnitudes fainter than H99, we are therefore searching for
stars with a much brighter radio-to-optical flux ratio.  However, a
FIRST---SDSS matching has the potential to reveal a radio-bright population too
rare to appear in the smaller H99 study, as the fainter SDSS includes a much
larger volume in the Galaxy.  For example: M dwarfs, many of which are known to
be magnetically active \citep[e.g.,][]{west08}, are found in significant
numbers only at faint magnitudes.  Thus, a search for radio-emitting M dwarfs
should be carried out in a deep optical survey with large sky coverage.
Advantages to using the SDSS are its high completeness, precise magnitudes, 
accurate astrometry, much deeper optical data than previous stellar catalogs,
and almost 300,000 stellar spectra.  

The remainder of the paper is laid out as follows.  In \S\ref{sec:surveys} we
describe the contributing surveys.  In \S\ref{sec:sample} we outline the
selection of the sample of radio stars and place an upper limit on the fraction
of radio stars.  In \S\ref{sec:optical} and \S\ref{sec:radio} we discuss
optical and radio properties of the sample, respectively, and in
\S\ref{sec:summary} we conclude and summarize our results.

\section{OPTICAL AND RADIO DATA}
\label{sec:surveys}

\subsection{Optical Catalog: SDSS}
\label{subsec:sdss}

We have drawn our sample from the photometric coverage of the sixth
data release (DR6) of the Sloan Digital Sky Survey\footnote{The survey website is
located at \emph{http://www.sdss.org}.} \citep[SDSS; see][and references
  therein]{york,edr,dr6}.  DR6 covers roughly $9,600\deg^2$ and contains 
photometric observations for 287\,million unique objects, as well as spectra
for more than 1\,million sources (in a smaller sky area of $6860\deg^2$).  SDSS
entered routine operations in 2000; DR6 observations were completed in June
2006.  Because SDSS spectroscopy is performed after photometric observations,
some stars in the sample have spectra that were not available until the Seventh
Data Release (DR7).  We have included some stars with DR7 spectra\footnote{We
  include stars with spectroscopic observations taken through 5 Dec 2007.}; we
point out explicitly where this detail affects our estimates of sample
contamination in \S\ref{subsec:contamination}.  The DR6 sample includes about
287,000 spectra classified as stars.

The SDSS photometric survey measures flux densities nearly simultaneously in
five wavelength bands ($u$, $g$, $r$, $i$, and $z$) with effective wavelengths
of 3551, 4686, 6165, 7481, and
8931\AA~\citep{fukugita,gunn,hoggPhoto,tucker,ugriz,ivezic04}.  Morphology
information allows reliable star---galaxy separation to $r\sim21.5$
\citep{lupton02,scranton}.  Sources are classified as resolved or unresolved 
using a measure of light concentration that determines how well the flux
resembles a point source \citep{edr}.  Magnitudes were corrected for Galactic
extinction according to the dust maps of \citet{sfd}.  The astrometry of the
photometric survey is good to $\lesssim0.1\arcsec$ \citep{sdss_astrometry}.

A subset of photometric sources was chosen for spectroscopy according to the
SDSS targeting pipeline; stars could be serendipitously selected by any of the
various targeting algorithms.  The quasar targeting algorithm selects all
$15<i<19.1$ sources within $2\arcsec$ of a FIRST catalog object, and some
sources as faint as $i=20.5$ (depending on the availability of spectral
fibers).  About 30\% of quasar targets turn out to be stars or galaxies
\citep{dr5quasars}.  Another algorithm targets interesting stellar classes by
selecting for their distinctive photometric colors; these include blue
horizontal-branch stars, carbon stars, subdwarfs, cataclysmic variables, brown
and red dwarfs, and white dwarfs.  Because stars in these catagories are
selected randomly to fill excess spectral fibers, the completeness of the
samples of stars with spectroscopic observations is, with a small number of
exceptions, not well-defined.  However, the fact that all objects within
$2\arcsec$ of a FIRST source are targeted implies that the spectroscopic sample
is complete to $i<19.1$ with respect to radio---optical sources brighter than
the FIRST limit, except in the case of fiber collisions (see below).

SDSS spectra are obtained using $3\arcsec$ fibers in a fiber-fed spectrograph
\citep{sdss_spectrograph}.  Spectra cover the wavelength range from about 3500
to 9500 \AA{} with a resolution of R$\sim$1800.  They are automatically
extracted and calibrated by the Spectroscopic Pipeline: wavelength calibrations 
are calculated from arc and night sky lines and flux calibrations from
observations of standard F stars.  Some fibers are used to observe blank sky
in order to correct targeted objects for the sky spectrum.  Not every object
targeted for spectroscopy obtains a spectrum, primarily because of fiber
collisions: due to the physical size of the fibers, no two fibers can be placed
closer than $55\arcsec$ on a spectroscopic plate.  As a result, spectra are
only taken for about 92.5\% of spectroscopic targets, although the completeness
effects are well understood \citep{tiling}.

\subsection{Radio Catalogs}

FIRST provides 20 cm radio fluxes, which were used to select candidate radio
stars.  The Westerbork Northern Sky Survey and the Green Bank 6 cm survey
provide fluxes at 92 cm and 6 cm, respectively, which allow us to determine a
radio spectral shape.

\subsubsection{FIRST}

The Faint Images of the Radio Sky at Twenty cm survey \citep[FIRST;][]{first}
used the Very Large Array to observe the sky at 20\,cm (1.4\,GHz) with a beam
size of 5\farcs4 and an rms sensitivity of about 0.15\,mJy~beam$^{-1}$.
Designed to cover the same region of sky as the SDSS, FIRST observed
$9,000\deg^2$ at the North Galactic Cap and a smaller $\sim2.5\degr$ wide strip
along the Celestial Equator from 1994 to 2002.   The survey contains over
800,000 unique sources, with positions determined to $\lesssim1\arcsec$; its
source density is roughly $97\deg^{-2}$.  It is 95\% complete to 2\,mJy and
80\% complete to the survey limit of 1\,mJy.  The integrated flux density,
$S_{20}$, is calculated using a two-dimensional Gaussian fit to each source
image. 

\subsubsection{WENSS}

The Westerbork Northern Sky Survey \citep[WENSS;][]{wenss} is a 92 cm radio
survey that was conducted in the mid-1990s.  It observed the sky
north of $\delta=29\degr$ to a limiting flux density of 18 mJy, with a beam size of
$54\arcsec$ and a positional uncertainty of $1.5$--$5\arcsec$.

\subsubsection{GB6}

The GB6 survey at 4.85\,GHz \citep[Green Bank 6\,cm survey;][]{gb6} was
executed with the original 91m Green Bank telescope in 1986 November and
1987 October.  Data from both epochs were assembled into a survey covering the
$0\degr<\delta<75\degr$ sky down to a limiting flux of 18\,mJy, with
$3.5\arcmin$ resolution and a positional uncertainty of $10$--$50\arcsec$.

\section{SELECTING A SAMPLE OF CANDIDATE RADIO STARS FROM FIRST AND SDSS}
\label{sec:sample}

\begin{figure}
\plotone{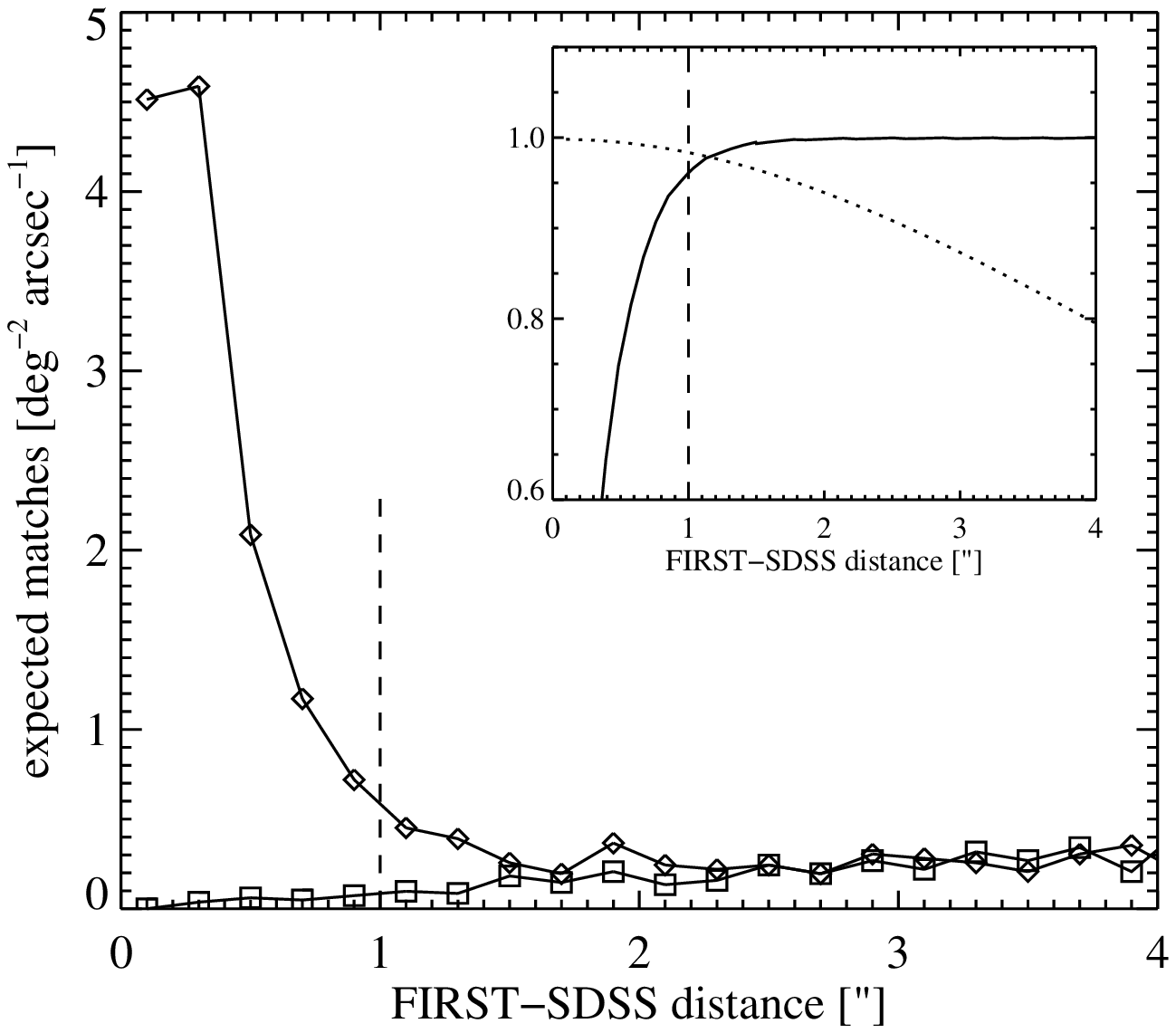}
\figcaption{\label{fig:astrometry}
The distribution of FIRST---SDSS distances for FIRST sources matched to SDSS
$r<21$ point sources (\emph{diamonds}).  The level of background contamination,
(estimated by offsetting positions by $1\degr$) is also shown (\emph{squares}). 
The inset plot shows the completeness (\emph{solid line}) and efficiency
(\emph{dotted line}) as a function of matching radius, estimated by fitting the
distance distribution with a simple Gaussian $+$ line model.}
\end{figure}

This section outlines the selection criteria that yield the candidate
radio stars sample.  We estimate the amount of contamination with random
matching, and use the result to place an upper limit on the fraction of radio
stars in the magnitude range of the sample.  The area of overlap of the two
surveys is about 9500 deg$^2$.

\subsection{Magnitude and flux limits}

We applied magnitude/flux limits as a way to control source quality.  For SDSS
sources, we required $r<20.5$ to ensure reliable determination of optical
morphology, as well as to select sources with a high enough signal-to-noise
ratio (S/N) for spectral typing.  Owing to the magnitude limits of the SDSS
quasar target selection algorithm (\S\ref{subsec:sdss}), most of the sources in
the final sample have $i\lesssim19.1$, although it is not an explicit
requirement.  There are roughly 4000--5000 SDSS sources with $r<20.5$ per
square degree on the sky, depending on Galactic latitude.

For the radio sources, we adopted a limit of $S_{20}\geq1.25$ mJy (equivalent
to AB magnitude 16.2).  This is slightly brighter than the 1 mJy depth of the
FIRST catalog; however, in a visual examination of faint FIRST images, many
sources fainter than this appeared to be possibly spurious detections.  Above
this flux limit, there are about 82 FIRST sources per square degree.

\subsection{Positional matching of point sources}

We restricted the sample to optical point sources using SDSS automated
star---galaxy separation (\S\ref{subsec:sdss}).  Figure~\ref{fig:astrometry}
shows the distribution of FIRST---SDSS distance for radio---optical (point
source) matches.  The expected contamination by random matches (evaluated by
off-setting the FIRST positions by $1\degr$ in right ascension) is also shown.
The inset plot shows the estimated completeness (solid line; percentage of
physical sources recovered) and efficiency (dotted line; percentage of all
matches which are physical) as a function of matching radius.  The precise
choice of matching radius is a tradeoff between sample completeness and
sample contamination.  Since we are looking for rare objects, we opted more on
the side of decreased contamination, and chose to use a $1\arcsec$ matching
radius.  As shown in Figure~\ref{fig:astrometry}, the matching radius of
$1\arcsec$ results in 96\% completeness and 98\% efficiency.
There are 2000--3000 SDSS point sources per square degree with $r<20.5$.
Correlating with FIRST positions within $1\arcsec$ resulted in $\sim14,000$
matches.

We did not consider proper motions when matching the two catalogs, but
emphasize that this decision should not significantly affect the matching
results.  All of the observations were performed since 1994.  Within the
relevant magnitude range, nearly 99\% of SDSS stars have an apparent motion of
less than 0.1 arcsec yr$^{-1}$, with a median value $<7$ mas yr$^{-1}$.  Proper
motions\footnote{Proper motions were determined from the Seventh Data Release
  of SDSS following the procedure of \citet{munn04}.} for this sample are much
smaller than for the sample of H99 owing to the much fainter flux limit and
correspondingly larger source distances.  

\subsection{Spectral typing}
\label{subsec:typing}

KI08 showed that a sample of potential radio stars selected by their SDSS
colors is strongly contaminated by quasars.  There are enough quasars with
stellar colors (i.e., within a few hundredths of a magnitude from the main
stellar locus in the multi-dimensional SDSS color space) that spectroscopic
identifications are necessary in order to cull them from the sample.  We
therefore required an SDSS spectroscopic observation for each matched source,
and limited the sample to those with stellar-classified spectra.  Of the
$\sim14,000$ selected matches, 6413 have spectra, and 292 of those were 
classified as stellar by the SDSS spectral processing pipeline.

For a more robust determination of spectral type, we executed visual
classification of each spectrum.  Of the 292 stellar-classified sources, 6
matched to known quasars \citep{dr5quasars} and 45 to known BL Lac objects
\citep{plotkin08}.  We removed one object whose spectrum showed broad emission
lines signifying the super-position of a star with a quasar.  Of course these
false matches are not indicative of an 18\% failure rate for the spectral
pipeline, which is optimized to detect galaxies, stars, and quasars rather than
rare sources such as BL Lac objects.  Because the selection criteria for this
study are biased toward (radio) quasars, they result in a disproportionately
large number of BL Lac objects, which make up a very tiny percentage of the
entire SDSS spectroscopic database.

Spectral types were individually assigned to the remaining 240 sources 
using a custom IDL package dubbed ``the
Hammer''.\footnote{\emph{http://www.cfa.harvard.edu/$\sim$kcovey/thehammer}}
The full algorithm used by the Hammer is described in Appendix~A of
\citet{covey2007}.  In short, the Hammer automatically types input spectra by
measuring a suite of spectral indices and performing a least-squares
minimization of the residuals between the indices of the target and those
measured from spectral type standards.  It then allows a confirmation or
correction of spectral type according to a \emph{visual} comparison of the
input spectrum with spectral templates.  Although spectral types are available
from the SDSS database, visual confirmation by stellar scientists (authors GRK,
AAW, JJB) leads to more robust classifications.

With visual classification, the sample was reduced to 194 sources with reliable
stellar spectra.  The rejected spectra were too noisy for reliable typing, or
were indicative of BL Lac objects.  The latter are not included in the BL Lac
sample of \citet{plotkin08}; that sample was drawn from SDSS Data Release Five
(DR5), and thus covers a smaller sky area than the DR6 sample used in this
paper.

\subsection{Visual examination of radio morphology}
\label{subsec:visual_radio}

We looked at FIRST images of the remaining sample in order to determine the
radio morphology of each source.  We anticipated that the majority would
be point sources, because the resolution of FIRST ($\sim5\arcsec$) is not
sufficient to resolve stellar emission.  Resolved or multiple-component
emission, however, would be strongly indicative of a non-stellar source such as
an active galactic nucleus (AGN) with radio jets.

\begin{figure}
\plotone{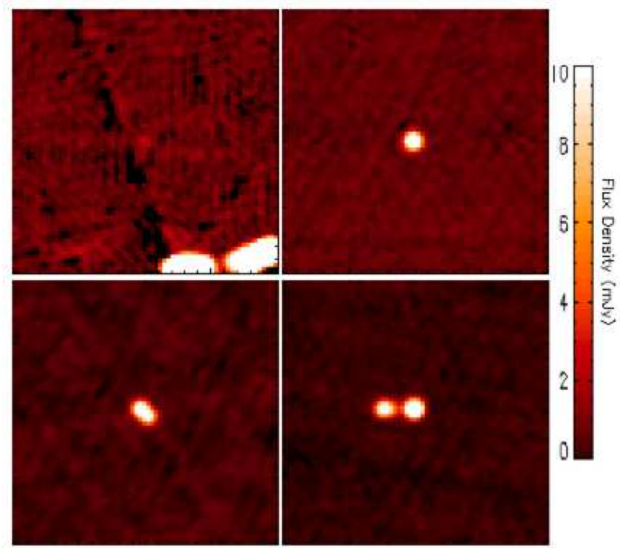}
\figcaption{\label{fig:morph_mosaic}
Example FIRST images ($2\arcmin$ by $2\arcmin$) classified by radio morphology
as spurious (\emph{upper left}), compact (\emph{upper right}), resolved
(\emph{lower left}), and complex (\emph{lower right}).  The position of the
optical source is at the center of each image.}
\end{figure}

We examined FIRST postage stamps ($2\arcmin\times2\arcmin$), and classified
each as ``compact'' (point source emission), ``resolved'' (resolved
single-component emission), ``complex'' (multiple-component or knotty
emission), or ``spurious'' (e.g., an artifact introduced by interferometric
errors).  An example from each category is shown in
Figure~\ref{fig:morph_mosaic}.  Resolved or complex radio morphology is 
typically associated with an extra-Galactic object such as a radio galaxy or
quasar.\footnote{Galactic objects such as HII regions or supernova remnants may
  show extended emission; however we do not expect contamination from these
  sources because our sample is limited to $30\degr$ or more above the Galactic
  plane.}  Out of the 194 visually-classified images we found six spurious 
sources, which were removed from the sample.  Of the remaining 188 sources,
60 are complex, 16 are resolved, and 112 are compact.\footnote{For one source
  that was initially identified as ``complex'', the optical image shows a
  nearby galaxy that appears to be the source of the second point of radio
  emission.  We concluded that the two radio components are physically
  unrelated, and moved that object into the ``compact'' category, leading to
  the totals given in the text.}  The high fraction of sources with complex
radio emission demonstrates that many radio quasars survived the previous
selection criteria.  Because these objects clearly show stellar spectra, we
interpret them as optically-faint radio quasars in chance alignment with bright
foreground stars.  We rejected the complex sources as obviously extra-Galactic.
However, not all quasars have complex radio emission from detectable lobes, but
many are instead point sources (KI08): this is typically thought to be the
result of Doppler beaming of a jet aligned along the line of sight.  Thus,
despite the thorough visual classifications, the final set of compact radio
sources with stellar spectra may remain contaminated by chance star---quasar
superpositions.  We discuss sample contamination in more detail in
\S\ref{subsec:contamination}.

\subsection{Final sample of potential radio stars}

We retain the sample of 112 compact sources for the remaining analysis of this
paper.  This set of radio star candidates is presented in
Table~\ref{table:radiostars}, which lists positions, stellar type, radio
fluxes, optical magnitudes, and radio morphology classification.  The 16
resolved sources and 60 complex sources are listed in
Tables~\ref{table:radiostars_resolved} and~\ref{table:radiostars_complex}, 
respectively.  The tables also list the distance to each star, calculated using
the photometric parallax relation of \citet{ivezic08}; this relation is valid
only for stars on the main sequence.  As discussed by \citet{finlator}, nearly
all SDSS stars ($\sim99\%$) are expected to lie on the main sequence.  However,
it is possible that by selecting radio-emitting stars we have biased our sample
toward giant stars in the halo (nearby giants are brighter than the $m=15$ SDSS
saturation limit).  We note that approximately 40\% of the candidate radio stars
have $\log(g)$ measured by the SDSS software pipeline; all have $\log(g)>3$,
indicating that they are main sequence stars.  There is no evidence that the
sample has a different distance distribution than other SDSS stars
with spectra.

\begin{figure}
\plotone{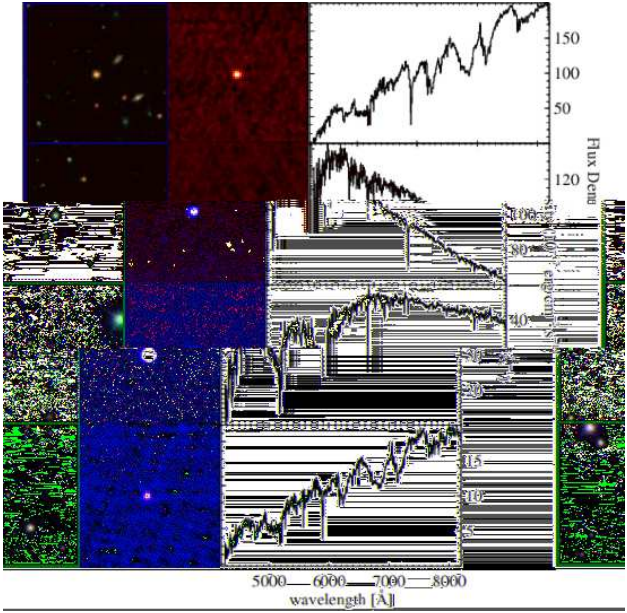}
\figcaption{\label{fig:star_mosaic}
Data for four of the FIRST---SDSS star matches.  The left column shows the
optical image ($2\arcmin\times2\arcmin$), a composite of the $g$, $r$, and $i$
band images; the middle column shows the FIRST image ($2\arcmin\times2\arcmin$)
with linear scaling; the right column shows the SDSS spectrum ($F_\lambda$).
The top row shows the brightest optical source, an M1 dwarf with $i=14.9$ and
$S_{20}=4.3$ mJy.  The second row shows a G3 dwarf with $i=16.0$ and
$S_{20}=17.1$ mJy.  The third row shows the brightest radio source, a K4 dwarf
with $i=16.7$ and $S_{20}=203$ mJy.  The bottom row shows an M1 dwarf with weak
H$\alpha$ emission (not visible on the scale of this figure), $i=17.8$, and
$S_{20}=6.6$ mJy.}
\end{figure}

Figure~\ref{fig:star_mosaic} is a mosaic of four of the
candidate radio sources; it shows the optical image, the radio image, and the
optical spectrum for each one.  The brightest optical source and the brightest
radio source in the sample are included in the mosaic.

\subsection{Estimating Sample Contamination}
\label{subsec:contamination}

To estimate the contamination originating from chance radio---optical
alignments, we created a set of twelve random samples for comparison.  To
create the random samples, we off-set the right ascension or declination in the
FIRST catalog (by $-1$, $-0.5$, $-0.1$, $0.1$, $0.5$, or $1$ degree), then
applied identical selection criteria (where possible) as above.  The
contamination estimate is an upper limit because it is not possible to apply
exactly the same selection criteria to the random samples: the SDSS spectral
target selection for quasars (\S\ref{subsec:sdss}) ensures that a large
fraction of the real FIRST---SDSS matches have spectral data, whereas very few
sources in the random samples have SDSS spectra.

Having performed the matching several times, we can determine the
variance of the random sampling.  \footnote{Varying the FIRST positions 
  results in a slight decrease of the areal overlap between FIRST and SDSS.
  However, given that so few matches result from the random sampling, the small
  change in matching area has no significant effect on the contamination
  estimate.}  Random matching of all FIRST and photometric SDSS sources
resulted in $3242\pm65$ matches.  Applying the $r$ magnitude and radio flux
limits reduced the samples to $759\pm30$ matches.  Selecting on optical point
source morphology further reduced the samples to $422\pm20$.

The next selection step for the real sample was to eliminate those sources
without SDSS spectra.  The equivalent step for the random samples is to
eliminate those which could not have qualified for SDSS spectral targeting.
The quasar targeting algorithm (\S\ref{subsec:sdss}) depends on proximity to a
radio source; by artificially shifting the FIRST positions to create the random
samples, we created fake optical---radio sources which would pass the selection
criteria.  We rejected sources outside of the DR6 spectroscopic coverage, which
is smaller than the DR6 photometric coverage.  Keeping only those matches which
would qualify for SDSS spectroscopy reduced the random samples to $225\pm22$.
Applying the success rate of spectral sampling of SDSS targets
\citep[92.5\%;][]{tiling} results in an expected random sample size of
$208\pm20$.

We can make an educated guess that the remaining random samples consist almost
entirely of stars: at these very bright magnitudes, the SDSS star---galaxy
separation mechanism is quite effective at differentiating between point
sources and extended sources.  Besides stars, the most common type of optical
point source is a quasar, but these are rare at $i<19.1$.  For example, the
highest fraction of quasars in SDSS can be found at the North Galactic Pole,
where there are about 100 stars for each quasar \citep[at magnitudes 
  $i<19$;][]{juric}.  Close to the Galactic plane, the quasar fraction is much
smaller.

The final step in the selection process was to visually classify each source
according to its radio morphology.  Only those objects which appeared by eye to 
be unresolved in their FIRST image were retained.  Selecting these objects
results in a random sample size of $117\pm14$.  Applying the spectral
observation success rate reduces that value to $110\pm14$.  Estimating one per
cent of the sample are quasars as described above, the final estimate is
$108\pm13$.  This number is essentially identical to the size of the candidate
radio stars sample, which consists of 104 stars from DR6 and 8 additional stars
with spectroscopy performed later than that of DR6.

The above comparison shows that most or all of the potential radio stars are
actually chance alignments of SDSS stars with unrelated FIRST sources.  The
variance in the random samples is large enough that there may be several real
radio stars in the candidates sample or none at all.  This result indicates
that radio stars are extremely rare or non-existant in the range $15<i<19.1$,
$S_{20}<1.25$ mJy.

\subsection{The Fraction of Radio Stars in the SDSS}
\label{subsubsec:fraction}

We can use the relative sizes of the candidate and random samples, along with
the sample completeness, to calculate an upper limit on the fraction of radio
stars in the SDSS.  The candidate sample contains 104 stars with spectra in
DR6; 98 of those are in the magnitude range $15<i<19.1$.  As discussed
previously, the estimated number of contaminating sources is $108\pm13$.  We
can therefore state with 97.5\% confidence that there are no more than 16 radio
stars in the sample of candidates (using the one-sided 2-$\sigma$ error
estimate).  The completeness estimate has the following contributions: 1) the
completeness of the FIRST survey at 1.25 mJy is approximately 85--90\%
\citep{first}; 2) the completeness of the radio---optical matching within
$1\arcsec$ is 96\%; 3) the SDSS spectroscopic targeting algorithm selects all
$15<i<19.1$ objects within $2\arcsec$ of a FIRST source (100\% completeness); 
4) given fiber collisions, the success rate of spectroscopic observations is
about 92.5\%\footnote{The success rate is not biased with respect to the radio
  sample; we verified that approximately 92.5\% of our selected sources which
  should have passed the targeting selection do in fact have SDSS spectra.}
\citep{tiling}.  We therefore estimate that the sample is about 75\% complete,
and thus that there are no more than 21 radio stars in the SDSS DR6 with
$15<i<19.1$.  There are approximately 18 million SDSS stars in this magnitude
range, which implies that no more than 1.2 out of every million stars in the
range $15<i<19.1$ have a radio flux of $S_{20}\geq$ 1.25 mJy.  For stars with
$i\approx15$, that corresponds to an upper limit on radio to optical flux ratio
of 0.34; for the $i\approx19.1$ stars at the faint end, the upper limit on
radio to optical flux ratio is 15.

\section{OPTICAL PROPERTIES}
\label{sec:optical}

We showed in the previous section that the sample is highly contaminated by
interloping AGN.  However, statistical comparisons of the sample with typical
stars can highlight the most likely actual radio stars.  In this section, we
examine optical properties of the sample: photometric colors, distance from the
stellar locus, spectral type, and magnetic activity.  We also use an \emph{SDSS
  control sample} selected from a strip of sky $1\degr$ wide in right
ascension.  The control sample contains point sources with $15<i<19.1$ and
$r<20.5$ from the region $236<\mathrm{R.A.}<237$, $-2.5\lesssim\mathrm{dec.}\lesssim65$;
it contains just over 160,000 sources.

\subsection{Photometric colors}
\label{subsec:colors}

\begin{figure}
\plotone{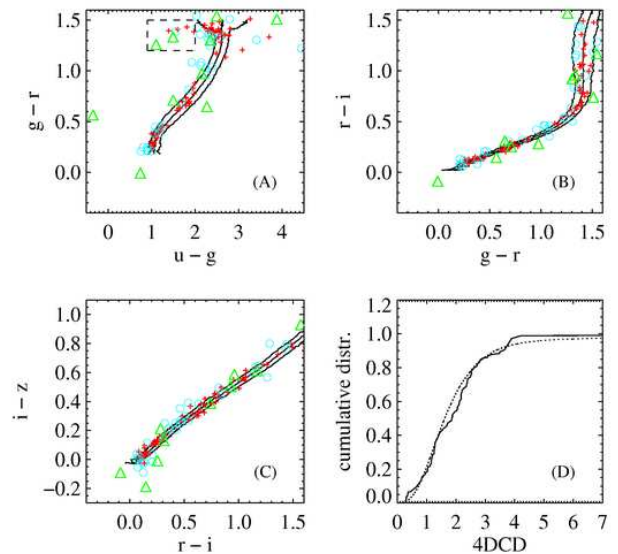}
\figcaption{\label{fig:chi}
Outliers from the stellar locus.  \emph{Panels A--C:} Color---color
diagrams with potential radio stars plotted as symbols.  The red plus signs
indicate candidate radio stars consistent with being on the
stellar locus, having 4DCD $<2$ (see text and Eq.~1 for definition); cyan
circles indicate those with $2<\mathrm{4DCD}<3$; green triangles indicate those
with $3<$ 4DCD.  The dashed-line rectangle in panel A surrounds the stars with
WD+dM bridge colors.  The SDSS stellar locus is shown by the solid lines, as 
parametrized by \citet{covey2007}.  The three lines indicate the position 
of the stellar locus and its interquartile width, projected onto two
dimensions.
\emph{Panel D:} Cumulative distributions of 4DCD for the candidate radio stars
sample (\emph{solid line}) and for the SDSS control sample (\emph{dotted
  line}).}
\end{figure}

Figure~\ref{fig:chi} presents the distribution of the candidate radio stars
in optical color---color space, compared to the SDSS stellar locus as
parameterized by \citet{covey2007}.  The majority of the candidate stars lie
on the stellar locus.  Several stars, however, appear to lie along the
white-dwarf---M dwarf (WD+dM) bridge \citep{smolcic}: $u-g<2, g-r>0.3,
r-i>0.7$.  Such close binary pairs are found to be more active than their
single field counterparts \citep[e.g.,][]{silvestri05,silvestri06}.
\citet{smolcic} limited their analysis to stars with $u<20.5$ to eliminate
those with poor photometry.  In the SDSS control sample, less than 0.1\% of
stars lie on the WD+dM bridge.  The candidate radio stars sample contains seven
stars which lie along the WD+dM bridge, corresponding to a much higher
fraction.  We note however that all of our WD+dM candidates have $u$-band
magnitudes $>20.5$; therefore $u$-band photometry may have artificially moved
these sources blueward of the M star locus (upper right corner of the stellar
locus).  None of their spectra suggest the presence of a white dwarf companion.
We note that the star with $u-g<0$ is not on the WD+dM bridge, as it has
$r-i\sim0.15$.  However, the spectrum and photometric image do suggest that it
is a (physical or optical) binary system: a K3 star with an added blue component.

A quantitative method of finding outliers from the stellar locus, taking
photometric errors into account, is outlined by \citet{covey2007}.  They
parameterized the stellar locus by finding its 1-$\sigma$ width in the standard
SDSS colors ($u-g, g-r, r-i, i-z$) and Two-Micron All Sky Survey \citep{2mass}
colors as a function of $g-i$.  They chose $g-i$ as the fiducial color because
it samples the largest wavelength range possible without relying on the
shallower $u$ or $z$ measurements.  Using only the SDSS colors, we define a
``four-dimensional color distance'' (4DCD; analogous to the seven-dimensional
version discussed in C07\footnote{Our definition of differs from C07 by a
  square root operator, such that our value is in the units of a color
  distance.}), which describes the statistical significance of the distance in
color space between a target object and the point on the stellar locus with the
same $g-i$ color as the target.  The 4DCD is defined by
\begin{equation}
\mathrm{4DCD} = \left[\sum_{k=0}^3 \frac{(X_k^\mathrm{targ}-X_k^\mathrm{locus})^2}{\sigma^2_X(\mathrm{locus})+\sigma^2_x}\right]^{1/2},
\end{equation}
where $X_0 = u-g$, $X_1 = g-r$, etc.  The width of the stellar locus is 
$\sigma_X(\mathrm{locus})$, which refers to the FWHM of the locus in color
$X_k$ at the same $g-i$ as the target object.  The error in the target's color,
$\sigma_x$, is calculated by adding in quadrature its photometric errors in the
two appropriate filters.

We characterize outliers from the stellar locus as those with a value of 4DCD
greater than two, shown as circles in Figure~\ref{fig:chi}.  Extreme outliers,
with 4DCD$>3$, are shown as triangles.  Three of the stars with WD+dM colors
have 4DCD$>2$; these are strong candidates for further investigation.  The
lower right panel compares the cumulative distributions for the candidate radio
stars (\emph{full line}) and the SDSS control sample; the two distributions are
similar.  The largest 4DCD value of $\sim14$ belongs to the possibly binary
star mentioned earlier, having $u-g<0$.  The other most extreme outlier has  
4DCD $\sim6.4$.  The spectrum of that object also shows some evidence that the
source is part of a multiple system, such as the super-position of a late-type
with an early-type M star.  All other values of 4DCD are $<4.3$.

\subsection{Spectral type and activity fraction}
\label{subsec:activity}

Because non-thermal radio emission is a signal of activity, we expect that
radio emission may correlate with strong spectral lines, such as H$\alpha$,
which are also known to signal magnetic activity.  We investigate the fraction
of active stars in our sample; if the sample contains some real radio stars
which are active, we may see an increase over the active fraction of \emph{all}
stars (without selecting for radio emission).  Previous studies have found that
the fraction of active M dwarfs is a strong function of spectral type
\citep{west04,west08}, tending to increase toward later subtypes with a peak
around M8 dwarfs.

\begin{figure}
\plotone{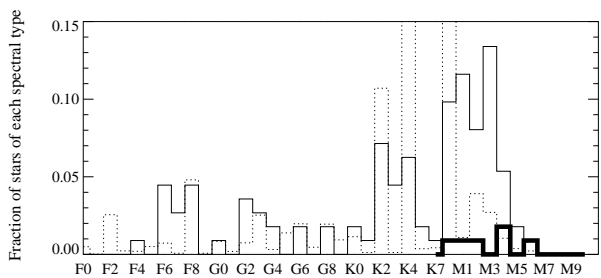}
\figcaption{\label{fig:activity}
Spectral types of the candidate radio stars (\emph{solid line}).  The thick
line indicates the six stars with H$\alpha$ in emission.  SDSS stars with
$15<i<19.1$ and $r<20.5$ are also shown (\emph{dotted line}).}
\end{figure}

An H$\alpha$ equivalent width was measured for each stellar spectrum using the
Hammer (\S\ref{subsec:typing}).  As discussed by \citet{west08}, the accuracy
of such measurements has been tested via Monte Carlo simulations to ascertain
how well line strength can be determined at a given S/N level.  The H$\alpha$
emission can be recovered over 96\% of the time for all spectral types.
Figure~\ref{fig:activity} shows the spectral types of the candidate radio stars
sample (\emph{solid line}), compared with all SDSS DR6 stars with spectra (just
under 1 million stars; \emph{dotted line}).  Spectral types for the SDSS sample
come from the automated version of the Hammer, while spectral types for the
candidate radio star sample were visually-confirmed (\S\ref{subsec:typing}).
Automated Hammer classifications are typically accurate to within $\pm4$
subtypes for A--G stars, and within $\pm2$ subtypes for K and M stars
\citep{covey2007}.  All spectra for the radio stars sample have S/N ratio
greater than 2.9.  As shown in Figure~\ref{fig:activity}, all of the active
candidate radio stars are M dwarfs.  This is not a
surprising result given that most stars are M dwarfs
\citep{covey08,bochanski09}, and the majority of activity in main sequence
stars is seen in M dwarfs \citep[e.g.,][]{gizis2002}.  \citet{west08} discussed
the fraction of active M dwarfs in SDSS DR5 as a function of spectral subtype,
after removing WD+dM bridge stars from their sample.  They found that M0--M3
stars have an active fraction of roughly 5--20\%, and that the fraction
increases strongly for M dwarfs of later subtype.  Our results for the
candidate radio stars are consistent with the results of that study for all
stars; we do not see a significantly higher fraction of active stars in the
sample of candidate stellar radio sources.

\begin{figure}
\plotone{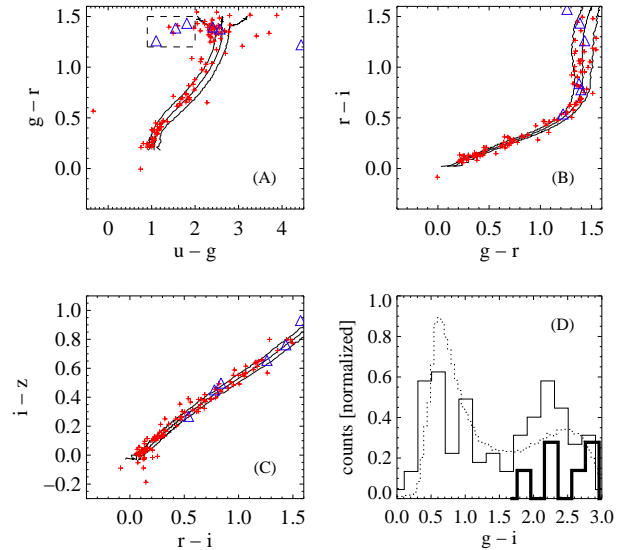}
\figcaption{\label{fig:colors}
Activity in the candidate radio stars sample.
\emph{Panels A--C:} Color---color diagrams with the inactive stars
indicated by red plus signs and the active stars (showing H$\alpha$ in
emission) indicated by blue triangles.  The dashed-line square in panel A
surrounds the stars with WD+dM bridge colors.  The SDSS stellar locus is shown
by the solid lines, as parametrized by \citet{covey2007}.  The three lines
indicate the position of the stellar locus and its interquartile width,
projected onto two dimensions.
\emph{Panel D:} The thin solid line shows the distribution of $g-i$ for the
candidate radio stars with $i<19.1$.  The thick line indicates the six 
active stars.  The dotted line corresponds to stars in the SDSS
control sample with photometric errors less than 0.1 mag 
in $g$ and $i$.} 
\end{figure}

Figure~\ref{fig:colors} shows color---color diagrams of the candidate radio
stars and the SDSS control sample, with the six active stars indicated by
triangles.  Three of the seven stars with WD+dM bridge colors
(\S\ref{subsec:colors}) are active; two of those are outliers from the stellar
locus (4DCD $>2$).  This result is in agreement with \citet{silvestri06}, who
suggest that 20-60\% of all WD+dM binaries are magnetically active.  The lower
right panel of Fig.~\ref{fig:colors} shows the $g-i$ distribution of the SDSS
control sample (\emph{dotted line}), the candidate radio stars (\emph{thin
  solid line}), and the active stars (\emph{thick solid line}).
\citet{covey2007} showed that $g-i$ correlates strongly with stellar spectral
type.  The distribution is bimodal in flux limited samples: it is biased toward
red stars, which are the most common, but also blue stars, which can be seen
from much greater distances.  \citet{west04} and \citet{bochanski07} compared
the colors of active to inactive M dwarfs in the field (i.e., not part of a
binary system).  They found no significant differences between the two
populations.  The current radio stars study includes three active field M
dwarfs (i.e., M dwarfs which do not have WD+dM bridge colors).  Although the
sample is too small to make a strong statement in comparison with the studies
of \citet{west04} and \citet{bochanski07}, we note that our results for stars
with radio emission are consistent with their conclusions for all stars.

\section{RADIO PROPERTIES}
\label{sec:radio}

\subsection{Radio Spectral Slope}

Using the multiple-wavelength radio catalog presented by KI08, we find that
eight of the candidate radio stars were detected in a second radio sky survey: 
either at 92 cm in WENSS or at 6 cm in GB6 (or both).  We can therefore
determine the radio spectral index $\alpha$ (where $F_\nu\propto\nu^\alpha$),
and we report those results here.  We compare the results with samples of the
type of AGN that may be contaminating the sample
(\S\ref{subsec:contamination}).

The value of the spectral index is a clue about the environment at the source
of emission.  Non-thermal radio emission is typically due to synchrotron
(relativistic) or gyrosynchrotron (semi-relativistic) electrons accelerating in
a magnetic field.  Synchrotron and gyrosynchrotron processes result in a
negative spectral slope ($\alpha\sim-0.8$) in the optically-thin case, and a
flat or positive slope in the optically-thick case.  Such emission has been
detected from M dwarfs in both their flaring (rapidly-variable) and quiescent
(non-flaring) states \citep[e.g.,][]{gudelbenz,large,bastian,osten06}.

\begin{figure}
\plotone{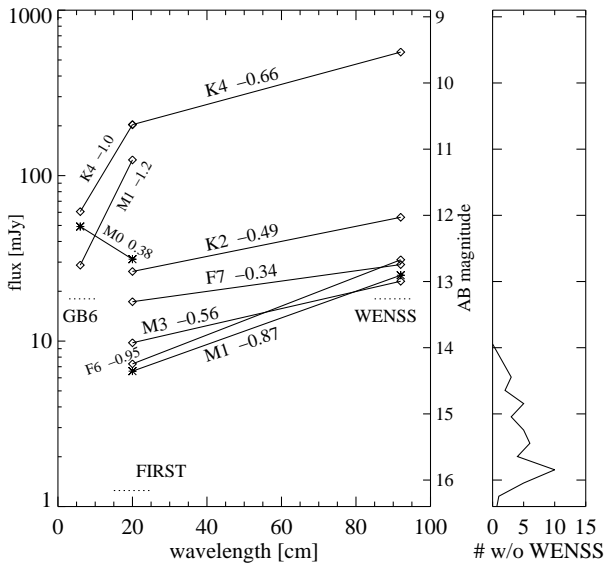}
\figcaption{\label{fig:wenss}
\emph{Left:} Radio spectra of the candidate radio stars detected at more than
one radio wavelength.  \emph{Diamonds} indicate inactive stars, \emph{asterisks}
indicate active stars.  The visually-confirmed spectral type and the power law
index are labeled for each source.  \emph{Right:} distribution of FIRST fluxes
for the 44 candidate radio stars which lie in the region of the sky observed by
WENSS, but were not detected at 92 cm.}
\end{figure}

Fifty stars in our sample lie within the WENSS sky coverage and 106 lie within
the GB6 sky coverage.  Following KI08, we use $30\arcsec$ as the FIRST---WENSS
matching radius and $70\arcsec$ as the FIRST---GB6 matching radius.  These
choices result in estimates of 99\% completeness with 92\% efficiency (WENSS),
and 98\% completeness with 79\% efficiency (GB6; see Table~2 of KI08).  Six
sources have 92 cm detections and three have 6 cm detections; only one source
was detected at all three wavelengths.  FIRST is much deeper than WENSS or GB6;
therefore only sources which are very bright or have very steep spectral slopes
can be detected in more than one survey.  Despite the poorer positional 
accuracy of GB6 and WENSS, these matches are highly reliable because of the
their lower source sky density.\footnote{For each object, we verified that the
  radio star candidate is the nearest FIRST source to the GB6 or WENSS match.}
Figure~\ref{fig:wenss} presents the radio spectra of these sources.  Four of
the stars are M dwarfs, two of which are active (\S\ref{subsec:activity}).

From the KI08 catalog, we selected a sample of AGN using the radio criteria
that were applied to the radio stars sample: $S_{20}\leq1.25$ mJy, unresolved
in FIRST (using Eq.~3 of KI08), and a WENSS match within $30\arcsec$.  These
criteria select the type of objects that could be contaminating the sample of
potential radio stars discussed in this section.  The stellar sample has a
similar spectral slope distribution to the AGN sample, whose median (mean)
spectral index is -0.67 (-0.54).  The spectral slope distributions do not allow
a  definitive conclusion as to whether any of the candidate radio stars with
multiple radio detections are actual radio stars or are contaminating AGN.

\subsection{Variability}

It is possible that the radio fluxes varied over the course of observations by
the different surveys.  Both stellar radio sources and AGN are known to vary in
the radio.  For example, \citet{berger02} observed variability in the
non-flaring radio emission from cool dwarfs over timescales of merely hours.
(However, the fluxes in that study were far below the flux limit of the current
study.)  The likely AGN contaminants are quasar core sources, which are known
to vary with timescales ranging from days to years
\citep[e.g.,][]{rys,barvainis}.  About half of the candidate radio stars sample
was observed muliple times in FIRST, most with timescales of less than a week.
None showed significant variability at 20 cm.  We note that nearly half of the
radio stars found by H99 varied in FIRST at the $4\sigma$ level.

\section{CONCLUSIONS}
\label{sec:summary}

We performed a search for radio stars by combining radio and optical
observations from FIRST (20 cm) and SDSS.  This is the first large-scale search
for radio stars using an optical survey faint enough to include a large number
of M dwarfs.  Many of these late-type stars are known to be magnetically
active.  A sample of 112 candidates was selecting using the following criteria:
optical point source morphology as determined by the SDSS photometric pipeline,
radio and 
optical positions matched within $1\arcsec$, an optical magnitude $r<20.5$ to
ensure reliable determination of optical morphology, radio flux
$S_{20}\geq1.25$ to eliminate spurious sources, a spectrum visually classified
as stellar, and radio point source morphology.  We estimated sample
contamination using random matching and similar selection criteria.  The size
of the random samples ($108\pm13$) suggests that the potential radio
stars are heavily contaminated by optically-faint radio quasars in chance
alignment with a foreground star.  The main ambiguity in determining
radio---optical matches stems from uncertainty as to whether the star is
actually the source of the radio emission.  It may be possible to overcome this
problem with careful follow-up using very long baseline interferometry to
determine proper motions of the radio sources.

In \S\ref{subsubsec:fraction}, we calculated the upper limit on the fraction of
radio stars (with $S_{20}\geq1.25$ mJy) no more than 1.2 per million stars in
the magnitude range $15<i<19.1$.  We note that some M stars have been observed
to occasionally flare brightly in the radio; however, these stars have a very
small duty cycle and very few will be detected in a single-epoch survey such as
FIRST.  While some M dwarfs have been shown to have constant radio emission
\citep[e.g.,][]{berger02,osten06}, it is at much fainter levels than radio
emission in their flare state by about an order of magnitude.  Our results
effectively rule out a population of radio bright late-type stars.

\begin{figure}
\plotone{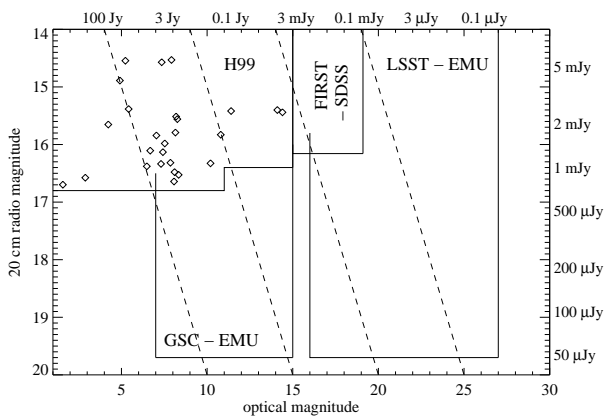}
\figcaption{\label{fig:parameterspace}
Parameter space probed by combinations of radio and optical surveys.  Stellar
radio sources detected by H99 are indicated by symbols.  Dashed lines show
constant radio-to-optical flux ratios of 1:1000, 1:100, 1:1, 100:1 (from left to
right).  The FIRST---SDSS matching corresponds to the current paper.  LSST and
EMU are future southern sky surveys.}
\end{figure}

We compared the candidate sample to SDSS stars in the same magnitude range and
investigated their distributions in optical color, stellar type, magnetic
activity, and distance from the stellar locus in color---color space.  The two
data sets show similar distributions.  However, there is a higher fraction of
stars with WD+dM bridge colors among the candidate sample.  Three of the radio
star candidates with WD+dM bridge colors are significantly offset from the
stellar locus in the four-dimensional color space (4DCD$>2$).  Two of those
three stars are magnetically active; a third star on the WD+dM bridge is also
active.  These four stars are good candidates for continuing investigation.

We searched for radio detections at two other wavelengths, 6 cm and 92 cm.
Eight sources were detected at more than one radio wavelength.  The spectral
index distribution for the stellar sources is similar to the distribution for a
sample of possible AGN contaminants.  Therefore the subset of stellar radio
candidates may be plagued by the same AGN contamination seen in the overall
sample.

This study shows that FIRST and SDSS are not a good pair of surveys for the
study (or discovery) of radio stars: stars bright enough at 20 cm to appear in
FIRST are probably above the $m=15$ saturation limit of the SDSS.
Figure~\ref{fig:parameterspace} shows the radio---optical parameter space
probed by the H99 study and the FIRST---SDSS correlation presented in this
paper.  FIRST is currently the largest deep radio survey available; the figure
indicates that much fainter radio data is required in order to discover more
stars of the population found by H99.  A much deeper sky survey at the same
frequency as FIRST is likely to be carried out in the southern hemisphere using
the Australian Square Kilometer Array
Pathfinder\footnote{\emph{http://www.atnf.csiro.au/projects/askap/}}
\citep[ASKAP;][]{askap}.  One of ASKAP's invited proposals involves a
project known as ``EMU: Evolutionary Map of the Universe'' (R. Norris et al. in
preparation), which includes a large southern sky survey down to 50 $\mu$Jy
($5\sigma_{rms}$) at 20 cm.  Although the main science driver for the ASKAP EMU
project is the study of AGN and galaxy evolution, the field of radio stars will
benefit immensely by cross-correlating the EMU survey with an all-sky optical
survey such as the Guide Star Catalog \citep[GSC;][]{gsc1} or a large southern sky
survey such as the Large Synoptic Survey
Telescope\footnote{\emph{http://www.lsst.org}} \citep[LSST;][]{lsst}.  

The GSC, the faintest of the optical surveys used in the H99 study, covered the
sky in the approximate magnitude range of 7-15.  As shown in
Fig.~\ref{fig:parameterspace}, the combination of the GSC and EMU will
be sensitive to stars with radio-to-optical flux ratios from less than 1:1000
to those with the highest flux ratios found by H99.  Covering half the sky and
extending 20 times fainter in the radio than FIRST, GSC---EMU should result in
the largest sample of candidate radio stars to date.

The LSST is a multi-epoch optical survey which will observe a quarter of the
sky every three nights and detect point sources down to $r\sim27$ ($r\sim24.5$
for a single exposure).  The radio---optical parameter space covered by a
potential LSST---EMU matching is also shown in Fig.~\ref{fig:parameterspace}.
The H99 survey found only a couple of radio stars of the type to which
LSST---EMU will be sensitive.  The EMU survey, 20 times fainter than FIRST,
will probe a 20--90 times larger volume in the Milky Way disk, and 
thus may find on the order of 100 new sources of faint stellar radio emission.
In addition, the LSST will be able to recognize active M dwarfs via their UV
flaring.  This method could be more efficient than looking for H$\alpha$
emission in SDSS spectra, which are available for only about 0.25\% of M dwarfs
detected by the SDSS.  The advent of the EMU survey to radio astronomy,
combined with large optical surveys, is likely to increase the number of known
radio stars by orders of magnitude.

\acknowledgements

This material is based upon work supported under a National Science Foundation
Graduate Research Fellowship, and by NSF grant AST-0507259 to the University of
Washington.  The authors would like to thank David Helfand and Nithyanandan
Thyagarajan for providing information about FIRST variability, and would also
like to thank Rachel Osten, Nicole Silvestri, Suzanne Hawley, Ray Norris, and
Jillian Bellovary for helpful discussions.  We also thank an anonymous referee
for suggestions that led to improvements in the content of the paper.

Funding for the SDSS and SDSS-II has been provided by the Alfred P. Sloan
Foundation, the Participating Institutions, the National Science Foundation,
the U.S. Department of Energy, the National Aeronautics and Space
Administration, the Japanese Monbukagakusho, the Max Planck Society, and the
Higher Education Funding Council for England. The SDSS Web Site is
http://www.sdss.org/. 

The SDSS is managed by the Astrophysical Research Consortium for the
Participating Institutions. The Participating Institutions are the American
Museum of Natural History, Astrophysical Institute Potsdam, University of
Basel, University of Cambridge, Case Western Reserve University, University of
Chicago, Drexel University, Fermilab, the Institute for Advanced Study, the
Japan Participation Group, Johns Hopkins University, the Joint Institute for
Nuclear Astrophysics, the Kavli Institute for Particle Astrophysics and
Cosmology, the Korean Scientist Group, the Chinese Academy of Sciences
(LAMOST), Los Alamos National Laboratory, the Max-Planck-Institute for
Astronomy (MPIA), the Max-Planck-Institute for Astrophysics (MPA), New Mexico
State University, Ohio State University, University of Pittsburgh, University
of Portsmouth, Princeton University, the United States Naval Observatory, and
the University of Washington.

\bibliography{bibliography}

\clearpage

\begin{deluxetable}{rrrrcrccrrrrrrrrr}
\tabletypesize{\scriptsize}
\rotate
\tablewidth{0pt}
\tablecaption{\label{table:radiostars}
Properties of the radio star candidates sample}
\tablehead{
\multicolumn{2}{c}{\emph {SDSS}\tablenotemark{a}} & \multicolumn{2}{c}{\emph
  {FIRST}\tablenotemark{a}} & 
\colhead{separation\tablenotemark{b}} & distance\tablenotemark{c} & & &
\colhead{$S_\mathrm{peak}$} & \colhead{$S_{20}$} & \multicolumn{5}{c}{SDSS
  model magnitudes} & & \colhead{catalog} \\ 
\colhead{R.A.} & \colhead{dec.} & \colhead{R.A.} & \colhead{dec.} &
\colhead{[$\arcsec$]} & \colhead{[pc]} &
\colhead{type\tablenotemark{d}} & \colhead{active?\tablenotemark{e}} &
  \colhead{[mJy/beam]} & 
\colhead{[mJy]} & \colhead{u} & \colhead{g} & \colhead{r} &
\colhead{i} & \colhead{z} & \colhead{4DCD\tablenotemark{f}} &
\colhead{ID\tablenotemark{g}} }
\startdata
   9.7469648 &   0.31168685 &     9.74689 &     0.31172 & 0.29 &    319.640 & M2 & no &       1.93 &       2.01 & 21.3 & 18.8 & 17.4 & 16.4 & 15.9 &  1.7 &      55105 \\
  27.6658730 &  -1.13948400 &    27.66569 &    -1.13928 & 0.99 &    397.880 & M2 & no &       1.82 &       1.96 & 21.6 & 18.9 & 17.5 & 16.6 & 16.2 &  1.8 &     155708 \\
  44.5010990 &   1.22593510 &    44.50115 &     1.22574 & 0.73 &   9830.540 & F8 & no &      11.90 &      12.40 & 19.2 & 18.3 & 18.1 & 18.0 & 18.0 &  2.8 &     253325 \\
 112.0061000 &  38.06249000 &   112.00596 &    38.06232 & 0.73 &    598.000 & M1 & no &       1.44 &       1.60 & 22.0 & 19.8 & 18.3 & 17.5 & 17.1 &  1.9 &     612859 \\
\hline
 114.5806600 &  48.39231500 &   114.58059 &    48.39211 & 0.76 &   2750.950 & K2 & no &       1.27 &       1.81 & 20.3 & 18.7 & 18.1 & 17.9 & 17.8 &  0.8 &     634578 \\
 117.5012900 &  34.98278400 &   117.50105 &    34.98291 & 0.85 &    691.950 & M1 & no &       1.83 &       1.56 & 21.3 & 18.9 & 17.6 & 17.0 & 16.6 &  2.5 &     660499 \\
 117.5527500 &  24.83105600 &   117.55283 &    24.83080 & 0.96 &   1139.890 & M0 & no &       2.97 &       2.85 & 23.1 & 19.7 & 18.4 & 17.9 & 17.6 &  2.2 &     661031 \\
 118.5574000 &  39.62222400 &   118.55721 &    39.62228 & 0.58 &   2268.550 & F8 & no &       1.27 &       1.29 & 16.9 & 15.9 & 15.6 & 15.5 & 15.5 &  0.4 &     670267 \\
\hline
 118.6814000 &  33.29291800 &   118.68152 &    33.29275 & 0.70 &    515.780 & M1 & no &       1.80 &       1.93 & 21.3 & 18.8 & 17.4 & 16.8 & 16.4 &  1.1 &     671427 \\
 119.3160100 &  30.17021500 &   119.31614 &    30.17024 & 0.41 &    936.480 & K4 & no &       3.57 &       3.82 & 19.8 & 17.7 & 16.7 & 16.4 & 16.2 &  3.5 &     677301 \\
 119.7809500 &  17.07585700 &   119.78081 &    17.07571 & 0.71 &   1613.270 & M0 & no &       2.37 &       1.79 & 23.2 & 20.2 & 18.9 & 18.4 & 18.1 &  1.2 &     681437 \\
 121.2057400 &  53.11331800 &   121.20609 &    53.11341 & 0.83 &   1070.120 & M3 & no &       3.81 &       4.97 & 22.6 & 21.1 & 19.8 & 18.8 & 18.3 &  3.7 &     694900 \\
\hline
 121.3575900 &  33.92661500 &   121.35767 &    33.92649 & 0.51 &   1040.680 & M3 & no &       1.74 &       1.25 & 23.6 & 21.6 & 20.2 & 19.2 & 18.6 &  1.2 &     696337 \\
 122.0831300 &  40.20902400 &   122.08297 &    40.20927 & 0.99 &  12403.900 & K0 & no &       1.48 &       1.36 & 20.2 & 19.2 & 18.9 & 18.8 & 18.8 &  1.2 &     703419 \\
 122.4434000 &  15.34121000 &   122.44323 &    15.34111 & 0.69 &    795.440 & M2 & ye &       1.45 &       3.43 & 22.8 & 20.2 & 18.9 & 18.0 & 17.5 &  1.9 &     706934 \\
 122.5364400 &  39.29135700 &   122.53619 &    39.29131 & 0.72 &    887.850 & K4 & no &       4.99 &       4.27 & 18.4 & 16.6 & 15.9 & 15.7 & 15.6 &  1.1 &     707887 \\
\hline
 123.5006500 &  29.96177800 &   123.50080 &    29.96161 & 0.77 &    535.690 & M4 &  m &       6.21 &       5.72 & 23.0 & 21.6 & 20.2 & 18.8 & 18.1 &  2.0 &     717543 \\
 123.5311000 &   7.57838460 &   123.53087 &     7.57825 & 0.94 &    395.760 & M4 & no &       1.01 &       1.58 & 23.6 & 20.8 & 19.3 & 18.1 & 17.4 &  2.4 &     717871 \\
 123.8783600 &  27.77808900 &   123.87866 &    27.77810 & 0.97 &   4003.250 & G2 & no &       2.84 &       2.45 & 19.5 & 18.3 & 17.9 & 17.8 & 17.7 &  1.9 &     721293 \\
 124.0573700 &  17.92303000 &   124.05750 &    17.92286 & 0.76 &   1131.870 & K4 & no &       1.38 &       1.88 & 19.2 & 17.4 & 16.7 & 16.4 & 16.3 &  1.2 &     723055 \\
\hline
 125.2091100 &  42.31718800 &   125.20885 &    42.31716 & 0.69 &    192.380 & M1 & no &       4.34 &       4.27 & 19.7 & 17.1 & 15.7 & 14.9 & 14.5 &  0.8 &     734833 \\
 126.3005300 &  17.33529000 &   126.30050 &    17.33502 & 0.98 &   1258.650 & M1 & no &       2.15 &       1.86 & 23.1 & 20.8 & 19.4 & 18.7 & 18.4 &  1.3 &     746124 \\
 127.3762300 &  47.77290800 &   127.37631 &    47.77266 & 0.92 &   1149.710 & K4 & no &     194.00 &     203.00 & 19.9 & 17.9 & 17.1 & 16.7 & 16.6 &  1.4 &     756772 \\
 128.4548800 &  28.86214000 &   128.45489 &    28.86231 & 0.61 &    315.370 & M3 & no &       9.40 &       9.77 & 22.1 & 19.7 & 18.4 & 17.1 & 16.5 &  2.4 &     767820 \\
\hline
 129.9784600 &  17.21355300 &   129.97861 &    17.21342 & 0.69 &    483.870 & M4 & ye &       2.50 &       2.46 & 23.2 & 21.6 & 20.2 & 18.8 & 18.0 &  2.4 &     783928 \\
 131.7995200 &   8.87368060 &   131.79954 &     8.87363 & 0.20 &   1409.130 & K3 & no &       1.67 &       1.89 & 19.9 & 17.9 & 17.2 & 16.9 & 16.8 &  1.2 &     803383 \\
 132.3854500 &  39.56273500 &   132.38580 &    39.56271 & 0.98 &    638.660 & M0 & no &       1.43 &       2.29 & 21.6 & 18.9 & 17.6 & 17.0 & 16.7 &  1.0 &     809871 \\
 133.7192300 &  22.38336800 &   133.71919 &    22.38349 & 0.46 &   3791.930 & F8 & no &       3.15 &       3.00 & 18.2 & 17.2 & 16.9 & 16.8 & 16.8 &  0.3 &     824171 \\
\hline
 133.9574700 &  16.77287300 &   133.95775 &    16.77288 & 0.95 &   2133.210 & G3 & no &      16.20 &      17.10 & 17.6 & 16.5 & 16.1 & 16.0 & 16.0 &  2.9 &     826748 \\
 138.3544300 &  16.92733700 &   138.35438 &    16.92730 & 0.22 &    930.940 & M2 & no &       4.84 &       4.56 & 22.5 & 20.4 & 19.1 & 18.3 & 17.8 &  2.6 &     874572 \\
 139.0118400 &  30.04233700 &   139.01158 &    30.04221 & 0.93 &    541.220 & M4 & ye &       2.16 &       2.39 & 23.2 & 21.4 & 19.9 & 18.7 & 18.0 &  1.3 &     881748 \\
 139.3560700 &  10.26608300 &   139.35585 &    10.26599 & 0.85 &   1857.790 & M0 & no &       5.43 &       5.64 & 23.1 & 20.7 & 19.4 & 18.9 & 18.5 &  1.6 &     885740 \\
\enddata 
\tablecomments{The table of 112 candidate radio stars is available at
  \emph{http://www.astro.washington.edu/users/akimball/radiocat/radiostars/}.} 
\tablenotetext{a}{Right ascension and declination are given in decimal
  degrees.}
\tablenotetext{b}{Offset between FIRST and SDSS positions.}
\tablenotetext{c}{Distance was determined using the photometric parallax
  relation of \citet{ivezic08}; the relation is valid only for stars on the
  main sequence.}
\tablenotetext{d}{Visually-confirmed spectral classification.}
\tablenotetext{e}{A ``yes'' indicates a spectrum with reliable H$\alpha$ emission.}
\tablenotetext{f}{Four-dimensional color distance from the stellar locus, defined in \S4.1 of the text.}
\tablenotetext{g}{Internal ID of the source in the radio catalog of KI08.}
\end{deluxetable}

\begin{deluxetable}{rrrrcrccrrrrrrrrr}
\tabletypesize{\scriptsize}
\tablewidth{0pt}
\rotate
\tablecaption{\label{table:radiostars_resolved}
``Candidate radio stars'' with resolved radio emission}
\tablehead{
\multicolumn{2}{c}{\emph {SDSS}\tablenotemark{a}} & \multicolumn{2}{c}{\emph
  {FIRST}\tablenotemark{a}} &
\colhead{separation\tablenotemark{b}} & distance\tablenotemark{c} & & &
\colhead{$S_\mathrm{peak}$} & \colhead{$S_{20}$} & \multicolumn{5}{c}{SDSS
  model magnitudes} & \colhead{catalog} \\ 
\colhead{R.A.} & \colhead{dec.} & \colhead{R.A.} & \colhead{dec.} &
\colhead{[$\arcsec$]} & \colhead{[pc]} &
\colhead{type\tablenotemark{d}} & \colhead{active?\tablenotemark{e}} &
\colhead{[mJy/beam]} & 
\colhead{[mJy]} & \colhead{u} & \colhead{g} & \colhead{r} &
\colhead{i} & \colhead{z} & \colhead{4DCD\tablenotemark{f}} &
\colhead{ID\tablenotemark{g}} }
\startdata
 123.1976100 &  20.64439200 &   123.19768 &    20.64418 & 0.80 &    351.420 & M1 & no &       1.25 &       2.11 & 20.6 & 18.0 & 16.6 & 15.9 & 15.5 &  0.6 &     714327 \\
 129.6339900 &  13.72514400 &   129.63398 &    13.72527 & 0.45 &   2147.920 & G4 & no &      16.60 &      30.00 & 18.4 & 17.2 & 16.7 & 16.6 & 16.5 &  4.6 &     780377 \\
 159.1123200 &  12.80009800 &   159.11230 &    12.80003 & 0.26 &  12229.900 & F7 & no &      68.30 &     103.00 & 20.2 & 19.3 & 19.1 & 19.0 & 18.9 &  3.5 &    1101138 \\
 179.5614700 &  37.07555200 &   179.56123 &    37.07574 & 0.97 &   6233.260 & K2 & no &       1.39 &       2.07 & 21.7 & 20.4 & 19.8 & 19.6 & 19.4 &  1.6 &    1324909 \\
\hline
 179.9687500 &  44.91827900 &   179.96864 &    44.91834 & 0.36 &   6285.120 & G5 & no &       3.73 &       6.41 & 20.4 & 19.5 & 19.1 & 18.9 & 18.8 &  3.1 &    1329461 \\
 183.0861100 &  27.54304500 &   183.08620 &    27.54281 & 0.89 &    687.640 & M2 & no &       4.66 &       7.81 & 23.2 & 20.3 & 18.8 & 18.0 & 17.5 &  0.8 &    1363672 \\
 189.3334100 &  53.88457500 &   189.33319 &    53.88464 & 0.52 &   2723.890 & G1 & no &       3.37 &       6.85 & 18.8 & 17.7 & 17.2 & 17.1 & 17.0 &  2.2 &    1431967 \\
 198.7216200 &  10.62265900 &   198.72157 &    10.62287 & 0.78 &   1404.190 & M2 & no &       3.75 &       7.46 & 22.8 & 21.2 & 19.9 & 19.1 & 18.4 &  5.6 &    1534251 \\
\hline
 217.8390300 &   8.53663970 &   217.83908 &     8.53640 & 0.88 &   2057.150 & K4 & no &       1.92 &       4.61 & 21.1 & 19.1 & 18.3 & 17.9 & 17.8 &  1.0 &    1742370 \\
 228.9030100 &   6.33068160 &   228.90314 &     6.33059 & 0.56 &   2877.940 & G1 & no &       2.90 &       5.90 & 18.3 & 17.4 & 17.0 & 16.9 & 16.8 &  3.0 &    1863654 \\
 240.3610400 &   9.07629920 &   240.36122 &     9.07639 & 0.72 &   1297.210 & M1 & no &      30.30 &      62.30 & 23.4 & 20.5 & 19.2 & 18.5 & 18.2 &  0.5 &    1985947 \\
 241.6174600 &  32.45298600 &   241.61725 &    32.45302 & 0.65 &    934.290 & M1 & no &      13.60 &      26.00 & 22.1 & 19.7 & 18.5 & 17.7 & 17.3 &  3.4 &    1998928 \\
\hline
 251.0205900 &  26.75291400 &   251.02068 &    26.75277 & 0.60 &   9124.150 & F8 & no &       3.67 &       7.14 & 19.9 & 18.9 & 18.7 & 18.5 & 18.5 &  2.8 &    2089662 \\
 253.7596800 &  32.12100800 &   253.75963 &    32.12106 & 0.24 &   1440.760 & K4 & no &       1.02 &       4.60 & 21.2 & 19.1 & 18.0 & 17.6 & 17.4 &  1.7 &    2113922 \\
 254.7470100 &  23.40834500 &   254.74695 &    23.40835 & 0.21 &    548.530 & M0 & ye &       1.27 &       2.70 & 20.7 & 18.6 & 17.3 & 16.6 & 16.2 &  3.9 &    2122609 \\
 258.2688900 &  32.22357000 &   258.26875 &    32.22338 & 0.81 &   2846.860 & F9 & no &       1.44 &       3.51 & 18.2 & 17.2 & 16.9 & 16.7 & 16.7 &  3.0 &    2152684 \\
\enddata
\tablecomments{The table of 16 sources with resolved radio emission (which
  passed all other selection criteria) is available at
  \emph{http://www.astro.washington.edu/users/akimball/radiocat/radiostars/}.}
\tablenotetext{a}{Right ascension and declination are given in decimal
  degrees.}
\tablenotetext{b}{Offset between FIRST and SDSS positions.}
\tablenotetext{c}{Distance was determined using the photometric parallax
  relation of \citet{ivezic08}; the relation is valid only for stars on the
  main sequence.}
\tablenotetext{d}{Visually-confirmed spectral classification.}
\tablenotetext{e}{A ``yes'' indicates a spectrum with reliable H$\alpha$ emission.}
\tablenotetext{f}{Four-dimensional color distance from the stellar locus, defined in \S4.1 of the text.}
\tablenotetext{g}{Internal ID of the source in the radio catalog of KI08.}
\end{deluxetable}

\begin{deluxetable}{rrrrcrccrrrrrrrrr}
\tabletypesize{\scriptsize}
\rotate
\tablewidth{0pt}
\tablecaption{\label{table:radiostars_complex}
``Candidate radio stars'' with complex radio emission}
\tablehead{
\multicolumn{2}{c}{\emph {SDSS}\tablenotemark{a}} & \multicolumn{2}{c}{\emph
  {FIRST}\tablenotemark{a}} & 
\colhead{separation\tablenotemark{b}} & distance\tablenotemark{c} & & &
\colhead{$S_\mathrm{peak}$} & \colhead{$S_{20}$} & \multicolumn{5}{c}{SDSS
  model magnitudes} & \colhead{catalog} \\ 
\colhead{R.A.} & \colhead{dec.} & \colhead{R.A.} & \colhead{dec.} &
\colhead{[$\arcsec$]} & \colhead{[pc]} &
\colhead{type\tablenotemark{d}} & \colhead{active?\tablenotemark{e}} &
  \colhead{[mJy/beam]} & 
\colhead{[mJy]} & \colhead{u} & \colhead{g} & \colhead{r} &
\colhead{i} & \colhead{z} & \colhead{4DCD\tablenotemark{f}} &
\colhead{ID\tablenotemark{g}} }
\startdata
   8.8779417 & -10.33142200 &     8.87798 &   -10.33148 & 0.25 &    365.070 & M3 & no &       4.87 &       5.79 & 22.6 & 19.7 & 18.3 & 17.2 & 16.6 &  0.7 &      50158 \\
  16.4108070 &   0.04541176 &    16.41080 &     0.04553 & 0.43 &   1605.120 & K7 & no &       1.36 &       4.61 & 22.9 & 20.3 & 19.0 & 18.5 & 18.3 &  3.2 &      92777 \\
  30.9980570 &  -9.00080110 &    30.99826 &    -9.00074 & 0.75 &   1409.850 & G2 & no &       1.41 &       1.97 & 16.7 & 15.6 & 15.2 & 15.1 & 15.1 &  0.4 &     174929 \\
  31.6274380 &  -8.36097310 &    31.62769 &    -8.36088 & 0.96 &   7284.830 & F8 & no &       1.05 &       1.96 & 20.0 & 19.0 & 18.6 & 18.5 & 18.5 &  1.3 &     178794 \\
\hline
 115.4349200 &  33.59706800 &   115.43505 &    33.59718 & 0.57 &    660.370 & M3 & no &     117.00 &     137.00 & 23.3 & 20.8 & 19.3 & 18.3 & 17.8 &  0.5 &     642007 \\
 116.5239800 &  33.14653100 &   116.52415 &    33.14652 & 0.51 &   1299.940 & G5 & no &      24.40 &      42.50 & 17.2 & 16.0 & 15.6 & 15.5 & 15.4 &  0.8 &     651724 \\
 118.6515700 &  31.04799200 &   118.65183 &    31.04788 & 0.89 &    976.060 & K2 & no &       3.14 &       5.70 & 18.2 & 16.6 & 15.9 & 15.7 & 15.6 &  1.1 &     671131 \\
 120.2777900 &  31.98149700 &   120.27754 &    31.98134 & 0.95 &   3123.470 & G5 & no &       1.40 &       1.80 & 18.7 & 17.5 & 17.2 & 17.0 & 17.0 &  1.1 &     686143 \\
\hline
 122.4168500 &   9.98404330 &   122.41704 &     9.98402 & 0.69 &   2606.500 & G8 & no &       9.55 &      22.70 & 19.4 & 18.1 & 17.6 & 17.4 & 17.3 &  1.1 &     706655 \\
 125.7096000 &  12.55254200 &   125.70952 &    12.55259 & 0.33 &    857.030 & M3 & ye &       7.58 &      15.40 & 23.0 & 21.4 & 19.9 & 19.0 & 18.5 &  4.2 &     740037 \\
 128.9825100 &  32.53530500 &   128.98271 &    32.53522 & 0.69 &    919.880 & F8 & no &      29.40 &      53.70 & 23.7 & 22.6 & 21.0 & 19.9 & 19.3 &  3.5 &     773185 \\
 131.5889500 &  13.51593600 &   131.58898 &    13.51595 & 0.10 &    881.950 & M3 &  m &       4.53 &       6.21 & 23.3 & 21.4 & 20.0 & 18.9 & 18.4 &  1.3 &     801062 \\
\hline
 131.9006900 &  15.72048400 &   131.90071 &    15.72021 & 0.99 &    916.410 & M3 & no &       1.08 &       3.20 & 22.6 & 21.5 & 20.0 & 19.1 & 18.6 &  3.7 &     804509 \\
 132.1393200 &   3.49329870 &   132.13913 &     3.49330 & 0.69 &   7584.010 & M2 & no &       5.51 &      13.10 & 20.3 & 19.4 & 19.0 & 18.9 & 18.9 &  2.7 &     807208 \\
 133.3787500 &   3.04040040 &   133.37881 &     3.04044 & 0.27 &   1017.230 & M1 & no &       5.67 &       8.47 & 22.9 & 20.4 & 19.1 & 18.3 & 17.9 &  0.7 &     820636 \\
 134.9379600 &  15.68712000 &   134.93808 &    15.68694 & 0.77 &   1733.890 & K4 & no &      29.40 &      32.10 & 21.0 & 19.0 & 18.1 & 17.7 & 17.6 &  1.6 &     837341 \\
\hline
 135.5234500 &  -0.30775603 &   135.52351 &    -0.30753 & 0.84 &   1238.090 & K7 & no &       2.59 &       3.77 & 21.8 & 19.4 & 18.2 & 17.7 & 17.5 &  1.1 &     843673 \\
 140.7074600 &   0.98319174 &   140.70767 &     0.98337 & 0.98 &   2584.960 & K2 & no &      10.50 &      16.70 & 19.6 & 18.3 & 17.7 & 17.5 & 17.4 &  1.6 &     900871 \\
 141.8337800 &  14.43029200 &   141.83377 &    14.43037 & 0.28 &    392.550 & M4 & no &      10.60 &      11.40 & 22.6 & 20.4 & 19.0 & 17.7 & 17.1 &  1.3 &     912857 \\
 147.9741100 &   0.86803458 &   147.97390 &     0.86804 & 0.76 &    426.410 & M4 & ye &       5.80 &       7.25 & 22.7 & 20.6 & 19.2 & 17.9 & 17.2 &  2.1 &     979415 \\
\hline
 148.9287500 &  40.24028700 &   148.92868 &    40.24017 & 0.46 &    301.060 & M4 & no &       2.90 &       5.75 & 22.6 & 20.2 & 18.8 & 17.4 & 16.7 &  1.2 &     989703 \\
 153.7401400 &  14.70054700 &   153.74037 &    14.70052 & 0.80 &   1375.330 & K7 & no &       7.83 &      11.60 & 21.7 & 19.2 & 18.1 & 17.6 & 17.4 &  1.6 &    1042751 \\
 154.0058700 &  12.36168900 &   154.00611 &    12.36181 & 0.95 &   2299.170 & K7 & no &       7.56 &       8.66 & 22.4 & 20.2 & 19.1 & 18.7 & 18.5 &  2.0 &    1045679 \\
 160.5310600 &   3.99054340 &   160.53126 &     3.99068 & 0.88 &    809.980 & M3 & no &       5.64 &       8.20 & 23.0 & 21.3 & 20.0 & 18.9 & 18.2 &  2.6 &    1116725 \\
\hline
 172.0590700 &  21.93210400 &   172.05921 &    21.93225 & 0.71 &    602.870 & M4 & no &       4.60 &       6.45 & 23.9 & 21.6 & 20.3 & 18.9 & 18.2 &  2.3 &    1242017 \\
 178.5105200 &   3.21151700 &   178.51063 &     3.21162 & 0.54 &  10917.700 & F4 & no &      54.30 &      72.20 & 19.9 & 19.0 & 18.7 & 18.7 & 18.7 &  1.1 &    1313182 \\
 186.8444200 &  46.53642400 &   186.84434 &    46.53632 & 0.42 &   2233.410 & M0 & no &       8.41 &      14.90 & 22.5 & 20.6 & 19.5 & 18.9 & 18.5 &  5.0 &    1404584 \\
 190.8415000 &  46.59573600 &   190.84160 &    46.59557 & 0.64 &   2809.610 & G5 & no &       1.06 &       3.81 & 18.9 & 17.8 & 17.3 & 17.2 & 17.1 &  2.6 &    1448337 \\
\enddata
\tablecomments{The table of 60 sources with complex radio emission (which
  passed all other selection criteria) is available in the electronic version
  of this paper, and is also downloadable from
  \emph{http://www.astro.washington.edu/users/akimball/radiocat/radiostars/}.}
\tablenotetext{a}{Right ascension and declination are given in decimal
  degrees.}
\tablenotetext{b}{Offset between FIRST and SDSS positions.}
\tablenotetext{c}{Distance was determined using the photometric parallax
  relation of \citet{ivezic08}; the relation is valid only for stars on the
  main sequence.}
\tablenotetext{d}{Visually-confirmed spectral classification.}
\tablenotetext{e}{A ``yes'' indicates a spectrum with reliable H$\alpha$ emission.}
\tablenotetext{f}{Four-dimensional color distance from the stellar locus, defined in \S4.1 of the text.}
\tablenotetext{g}{Internal ID of the source in the radio catalog of KI08.}
\end{deluxetable}

\end{document}